# Maintaining Capabilities in CCD Production for the Astronomy Community

State of the Profession Considerations: Instrumentation and CCD Fabrication

Kyle Dawson (University of Utah): kdawson@astro.utah.edu
Stephen Holland (Lawrence Berkeley National Laboratory)
David Schlegel (Lawrence Berkeley National Laboratory)

CCD detectors play a vital role in all aspects of optical astronomy.  Critical to advancing research is the ability to partner with commercial foundries to produce custom devices that meet the needs of specific instruments.  For more than 20 years, Teledyne DALSA Semiconductor was the primary industrial partner in the manufacturing of 150 mm wafers for CCDs.  DALSA is migrating the manufacturing from 150mm to 200mm wafer diameter and will not be updating their CCD processing tools for the new format wafer.  As a result, DALSA will no longer serve as a partner to the astronomy community in the manufacturing of CCDs.  We recommend that the Department of Energy, National Science Foundation, and NASA jointly pursue a new commercial partner in CCD fabrication to maintain capabilities in custom CCD design for astronomy applications.

**Key Issue and Overview of Impact on the Field**

CCD development for CCDs with enhanced quantum efficiency at wavelengths up to one micron was performed at the Lawrence Berkeley National Laboratory (LBNL) in partnership with DALSA Semiconductor. Over a twenty-year period, these fully depleted CCDs advanced from an early prototype to a device that can be manufactured in bulk with few cosmetic defects and nearly optimal sensitivity across the wavelength range $3{,}600 < \lambda < 10{,}000$ Å. These detectors are the enabling technology in the DES, BOSS, and eBOSS dark energy experiments and are being instrumented in the DESI spectrographs.

This fully depleted technology has now been licensed by Hamamatsu and used to fabricate fully depleted CCDs for the Subaru Hyper-Suprimecam imager; STA and Teledyne/e2V have also used this technology to fabricate CCDs for LSST.  With these efforts, Silicon CCDs with conventional readout sequences have matured for optical bands covering $3{,}600<\lambda<10{,}000$ Å, thus demonstrating the advantages of a commercial partnership to develop new technologies with specific astronomy applications.  Similar developments have occurred within NASA and other instrumentation labs across the country.  While Hamamatsu and Teledyne/e2V will be able to sustain production of conventional astronomy CCDs, there is not a clear mechanism to partner with these vendors to advance new technologies.  With the departure of DALSA, there is not a clear partner to maintain R&D activities in CCD technology.

DOE laboratories are exploring new 200 mm foundries to maintain R&D capabilities for future CCD design and fabrication.  Several foundries specializing in CMOS image sensors (CIS) and one that has in place a traditional triple-polysilicon CCD process appear to be possible partners.  A transition to a new vendor offers opportunities to produce larger format Silicon detectors with enhanced capabilities.  New Silicon detectors such as lower noise CCDs should be achievable with the more advanced CIS processes.

More importantly, by developing a new path toward fabrication and testing of CCDs, the astronomy community will maintain the ability to develop customized detector formats. Two examples of new detector possibilities are as follows:

**Germanium CCDs:** the effective band-gap around 1 eV limits the effectiveness of Silicon CCDs at redder wavelengths. The primary spectroscopic feature used to determine redshift in galaxy surveys is due to [OII] emission which appears beyond the 10,000 Å cutoff in a silicon detector for galaxies at redshifts $z > 1.6$. Enormous, relatively unexplored volumes will still be available at these higher redshifts even after DESI is completed. Many cosmological models (such as early dark energy) are best explored by measuring the expansion history and growth rate at these high redshifts. Spectroscopic observations of galaxies and clusters at epochs near the peak of star formation can inform models of galaxy formation while also gaining access to the cosmic web at high redshift. Detectors with sensitivity at wavelengths longer than 10,000 Å allow us to extend galaxy surveys and spatially resolved spectroscopy to these higher redshifts.

While infrared InGaAs and HgCdTe CMOS detectors have been used in ground- and space-based observatories, these detectors are expensive, require substantial cooling, and suffer from low yield in the fabrication process. Calibration of the pixelated amplifiers is a considerable challenge and an impediment to data quality for long term surveys. Germanium CCDs offer a potentially attractive alternative to access longer wavelengths. Germanium CCDs can be processed with the same tools used to build silicon imaging devices, show promise for read noise and sensitivity comparable to that of Silicon detectors, and offer a high quantum efficiency to wavelengths as red as 1.4 microns when cooled to 77 K (Figure 1). This increase in wavelength coverage will allow a spectroscopic identification of [OII] emission lines to $z = 2.6$. The manufacture of Germanium CCDs is a current focus of research and development at DOE labs and at Lincoln Labs. It is likely that Germanium CCD manufacturing will require substrates that are compatible with a silicon foundry, such as bonded Si-Ge wafers or a Ge on insulator on silicon substrate. In any case, the effort requires the new partnership with a CCD foundry to advance.

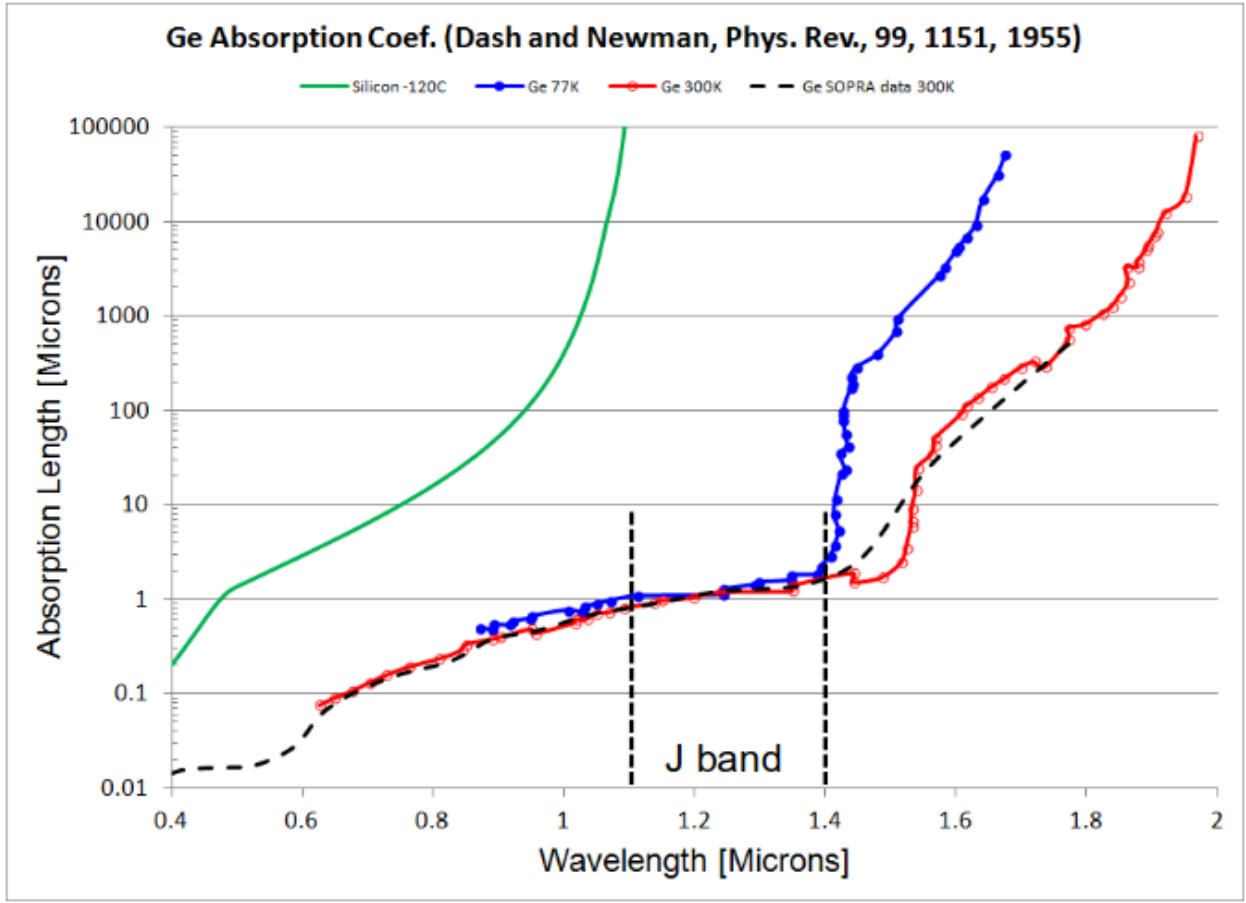

**Figure 1:** Absorption length as a function of wavelength for Silicon (green), Germanium cooled to 77 k (blue), and Germanium at room temperature (red). Fabrication of Germanium CCDs faces several challenges that need to be addressed before these devices can be integrated onto large focal planes. Several processes in doping, etching, and film deposition are similar to those in silicon CCD fabrication, but need to be tested with a new vendor's capabilities.

**Skipper CCDs on Silicon with single-photon sensitivity:** a Skipper CCD design on Silicon (https://arxiv.org/abs/1706.00028) with ultra-low readout noise shows promise for single photon detection. Several methods for direct detection of dark matter using Silicon detectors would benefit from the increased sensitivity to rare events offered by these new CCDs. Low-noise, visible detectors can satisfy the needs for faint exoplanet characterization with a space-based instrument. Detectors with minimal readout noise would also enhance the sensitivity of high-resolution spectrographs or spectroscopic observations in the far blue that would otherwise be readnoise limited.

    The Fossum group at Dartmouth has reported so-called "Golden pixels" in CMOS image sensors that achieve ~0.2 e- RMS noise in a single read (https://www.mdpi.com/1424-8220/16/8/1260). Incorporating that capability into Skipper

CCDs would greatly reduce the readout time that is the major impediment for the use of Skipper CCDs in astronomy applications. The CIS foundry approach also offers the opportunity for integration of CMOS circuitry on the same chip as the CCD. This will require experienced CMOS design engineers and close partnership with a commercial foundry.

## Strategic Plan

There are two key milestones to qualify a new commercial partner for CCD fabrication.

The first step is to procure 200 mm high-resistivity silicon and qualify the wafers for the 200 mm foundries. In collaboration with the foundries, the detector R&D community must develop gettering methods for the removal of harmful impurities in the silicon. We need to fabricate full thickness pin diodes on high-resistivity silicon wafers in a CIS foundry and a CCD foundry along with small format CCDs that are compatible with the die size limitations of wafer stepper photolithography. The results will be used to demonstrate low dark current on full thickness, i.e. 725 micron-thick, fully depleted devices. The CIS CCD in CMOS will be compared to conventional triple-polysilicon CCDs.

The second step is to fabricate large format CCDs at the CIS foundry and the CCD foundry via development of photolithography stitching methods. It will be crucial to compare yield and performance between the two approaches especially as relates to intra-level shorts. We note that the CCDs in CMOS processes are significantly more susceptible to intra-level shorts than in the triple-polysilicon process. It must then be determined if the more advanced CCD in CMOS technology, e.g. ~ 180 nm versus 2-3 um for the CCD foundry, can overcome the intra-level shorts issue. Finally, we must produce thinned, back-illuminated CCDs using CIS industry-standard methods of wafer bonding, thinning, ion implantation, and laser annealing.

## Organization, Partnerships, and Current Status

Sustaining CCD capabilities is not only essential for optical astronomy but also synergizes with NASA x-ray astronomy, light source applications at DOE labs, and direct searches for dark matter. The first of those two applications would benefit from the higher quantum efficiency for 10-30 eV x-rays in Ge relative to silicon, thus extending sensitivity to harder x-ray sources. There are also relevant applications in accelerator research, quantum information science, and nuclear security. A separate white paper requesting support for detector R&D has been submitted to the DOE-HEP cosmic frontiers program, but there is no clear path to financially support a full exploration of a new commercial vendor.  Potential exists to partner in detector R&D across several disciplines.

## Schedule

We estimate that two years are required to fully vet a commercial partner for CCD fabrication.  The time may be shortened if testing is performed across multiple labs with support from multiple agencies.

## Cost Estimates

The costs for the 200 mm foundry runs are not yet established, but we expect that they will be substantially higher than the costs at Teledyne DALSA, which were about ~$150 to $200k for a foundry run with a new mask set.  Two runs are expected at the CIS foundry and two runs are expected at the CCD foundry.

The risk is lower for the foundry with the established CCD process, but given the 2 - 3um design rules there is less possibility of performance improvement.  The CIS foundries can produce transistors an order of magnitude smaller in size, and the reduction in capacitance would result in lower noise performance.